\documentclass[%
reprint,
 showpacs,
 amsmath,amssymb,
]{revtex4-1}
\usepackage{graphicx}% Include figure files
\usepackage{dcolumn}% Align table columns on decimal point
\usepackage{bm}% bold math
\usepackage{CJK}
\usepackage{color}
\usepackage{amssymb}
\usepackage{amsfonts}

\definecolor{Red}{rgb}{1,0,0}

\usepackage[colorlinks,urlcolor=blue,linkcolor=blue,citecolor=blue,anchorcolor=blue]{hyperref}
\makeatletter

\newcommand{\Rmnum}[1]{\expandafter\@slowromancap\romannumeral #1@}
\makeatother
\begin{document}
%\begin{CJK}{GBK}{song}

\preprint{APS/123-QED}

\title{Triphoton generation near atomic resonance via SSWM: Harmonic expansion for accurate optical response}

\author{Jianming Wen*}
\affiliation{Department of Electrical and Computer Engineering, Binghamton University, Binghamton, New York 13902, USA}
\email{jwen7@binghamton.edu}

\date{\today}% It is always \today, today,

\begin{abstract}
\noindent Quantum correlations of time-frequency-entangled photon pairs generated via parametric processes are critically influenced by both the linear and nonlinear optical responses of the medium. This sensitivity is especially significant in schemes utilizing atomic ensembles with well-defined energy level structures near resonance. However, conventional theoretical approaches often fall short in accurately calculating the optical responses--particularly when a single atomic transition is simultaneously driven by multiple light fields with (significantly) different intensities. To address this limitation, we generalize the harmonic expansion method originally introduced by Wen for biphoton generation near atomic resonance. As a case study, we apply this generalized approach to the reliable direct generation of time-energy-entangled W-state triphotons via spontaneous six-wave mixing in a five-level asymmetric-M atomic system. Our results demonstrate the method's superior accuracy and self-consistency, offering clear advantages over traditional calculation techniques.

\end{abstract}

%\pacs{42.25.--p, 42.82.Et}% PACS, the Physics and Astronomy
\maketitle

\emph{Introduction}---Time-energy-entangled photon pairs play a crucial role in a wide range of quantum technologies, including quantum communication and networking, where their strong temporal correlations enable secure and efficient information transfer. They are also essential in quantum metrology for applications such as high-precision timing and synchronization, and serve as reliable carriers of entanglement in quantum computing and information processing. Owing to their robustness against dispersion, these photon pairs are well-suited for long-distance transmission in optical fibers. Moreover, their ability to operate at low light levels with high temporal resolution makes them highly valuable for quantum spectroscopy in biology, where they allow for non-invasive probing of fragile biological samples.

These entangled photon pairs are typically generated through light-matter interactions in nonlinear optical materials, and their correlation properties are intrinsically determined by the optical response, including both linear and nonlinear effects. This dependence is especially significant in systems employing atomic ensembles with well-defined energy-level structures \cite{1,2,3,4,5,6,7,8,9,10,11,12,13,14,15,16,17,18,19,20}, where the interaction between atomic transitions and driving fields governs not only generation efficiency but also the temporal coherence and entanglement characteristics of the emitted photons. Accurately modeling the optical response in such systems remains a central theoretical challenge, particularly when a single atomic transition is simultaneously driven by multiple light fields with (substantially) different intensities or detunings \cite{21,22,23}. Such conditions, common in (spontaneous) multi-wave mixing processes, can render standard perturbative approaches \cite{24,25,26,27,28,29,30} inadequate, often leading to oversimplified or inconsistent results. 

To overcome these limitations, we extend the harmonic expansion (HE) approach originally proposed by Wen \cite{21,22,23} for narrowband biphoton generation through spontaneous four-wave mixing (SFWM) near atomic resonance, generalizing it to accommodate more complex interaction schemes. As a representative example, we apply this method to the direct generation of time-energy-entangled W-class triphotons via spontaneous six-wave mixing (SSWM) in a five-level asymmetric-M atomic system. This process, involving multiple resonant and near-resonant excitation pathways, provides an ideal testbed for assessing the accuracy and flexibility of the generalized approach. Notably, SSWM has recently emerged as the only viable mechanism for the direct generation of genuine and reliable W-triphotons in coherent atoms \cite{30}. Our theoretical framework captures the relevant physical mechanisms with greater precision and resolves key ambiguities encountered in conventional perturbative treatments. The resulting predictions for both the optical response and the three-photon time correlations confirm the method's enhanced accuracy and internal consistency, establishing it a powerful tool for analyzing and designing advanced quantum optical systems.

\begin{figure}[t]
\centering
  \includegraphics[width=7cm]{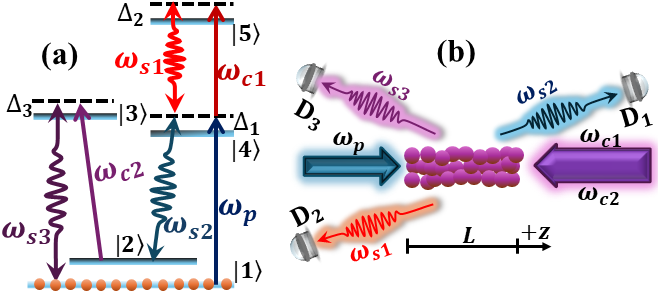}  %\columnwidth
  \caption{Direct generation of continuous-mode, time-energy entangled W-triphotons via single-step SSWM in a laser-cooled five-level atomic system with an asymmetric-M structure. (a) Energy-level scheme: one weak pump field E$_p(\mathbf{k}_p,\omega_p)$ and two strong coupling fields E$_{c1}(\mathbf{k}_{c1},\omega_{c1})$ and E$_{c2}(\mathbf{k}_{c2},\omega_{c2})$ drive the system, leading to spontaneous W-triphoton simultaneous emission $\hat{E}_{s1}(\mathbf{k}_{s1},\omega_{s1})$, $\hat{E}_{s2}(\mathbf{k}_{s2},\omega_{s2})$, and $\hat{E}_{s3}(\mathbf{k}_{s3},\omega_{s3})$ through SSWM. (b) Spatial configuration of incident lasers and emitted triphotons.}
  \label{fig1}
\end{figure}

\emph{Model}---The five-level asymmetric-M atomic system is schematic in Fig.~\ref{fig1}(a), where cold atoms are confined within a thin cylindrical volume of length $L$ and density $N$. The ensemble is initially prepared in the ground state $|1\rangle$. The geometrical arrangement of input and generated fields in the SSWM process is illustrated in Fig.~\ref{fig1}(b). A weak pump field E$_p$ with wave vector $\mathbf{k}_p$ and frequency $\omega_p$ and two strong coupling fields E$_{c1}(\mathbf{k}_{c1},\omega_{c1})$ and E$_{c2}(\mathbf{k}_{c2},\omega_{c2})$ counter-propagate along the $z$-axis through the ensemble. Specifically, E$_p$ drives the $|1\rangle\rightarrow|4\rangle$ transition with a large blue detuning $\Delta_p=\omega_{41}-\omega_p$; E$_{c1}$ couples $|4\rangle\rightarrow|5\rangle$ with detuning $\Delta_{c1}=\omega_{54}-\omega_{c2}$; and E$_{c2}$ couples $|2\rangle\rightarrow|3\rangle$ near resonance with detuning $\Delta_{c2}=\omega_{32}-\omega_{c2}$, where $\omega_{ij}$ is the transition frequency from $|j\rangle$ to $|i\rangle$. The fifth-order nonlinear susceptibility $\chi^{(5)}$ mediates phase-matched production of narrowband W-triphotons $\hat{E}_{s1}(\omega_{s1},\mathbf{k}_{s1})$, $\hat{E}_{s2}(\omega_{s2},\mathbf{k}_{s2})$, and $\hat{E}_{s3}(\omega_{s3},\mathbf{k}_{s3})$, which are spontaneously emitted under simultaneous energy and momentum conservation. In this setup, the strong field E$_{c2}$ and the generated $\hat{E}_{s3}$ photons form a standard three-level $\Lambda$-type electromagnetically induced transparency (EIT) system \cite{31}. As a result, E$_{c2}$ not only enhances the SSWM process but also opens a transparent window for $\hat{E}_{s3}$ with a pronounced slow-light effect. This EIT-assisted interaction brings the $\hat{E}_{s3}$ photons close to resonance, significantly boosting the overall efficiency of triphoton generation.

In theory, the SSWM process \cite{27,28,29,30} is governed by the effective interaction Hamiltonian
\begin{eqnarray}\label{eq1}
H=\epsilon_0\int_VdV\chi^{(5)}\text{E}_1\text{E}_2\text{E}_3E^{(-)}_{s1}E^{(-)}_{s2}E^{(-)}_{s3}+H.c.,
\end{eqnarray}
where the three input (generated) fields are treated as classical (quantized) and $V$ is the interaction volume. In the Schr\"{o}dinger picture, applying first-order perturbation theory and neglecting the pure vacuum contribution, the resulting true triphoton state vector at the output surfaces is
\begin{eqnarray}
|\Psi\rangle = \Psi_0\iiint d\omega_{s1}d\omega_{s2}d\omega_{s3}\chi^{(5)}\Phi\Big(\frac{\Delta kL}{2}\Big)\delta(\Delta\omega)|111\rangle,  %|1_{\omega_{s1}},1_{\omega_{s2}},1_{\omega_{s3}}\rangle,
\end{eqnarray}
where $\Psi_0$ is a normalized constant, $\Delta\omega=\sum_{j=1}^3(\omega_j-\omega_{sj})$, and $\Delta k=\sum^3_{j=1}(\mathbf{k}_j-\mathbf{k}_{sj})\cdot\hat{z}$. The Dirac $\delta$-function $\delta(\Delta\omega)$ enforces strict energy conservation, ensuring that the atomic population cycle closes and returns to its initial state. The spectral and temporal characteristics of $|\Psi\rangle$ are dictated by the combined effects of the complex phase-matching function $\Phi(x)=\text{sinc}(x)e^{-ix}$ and $\chi^{(5)}$. The function $\Phi$ encapsulates the medium's linear optical response and sets the natural spectral width through group-velocity mismatch and transmission filtering. By contrast, $\chi^{(5)}$ governs the generation mechanism and rate, and defines the intrinsic spectral profile and bandwidth of the newborn triphotons. 

These spectral features manifest directly in correlation measurements, with the effective coincidence window set by the narrower of the bandwidths. Consequently, two distinct temporal-correlation regimes emerge, dominated respectively by $\Phi$ and $\chi^{(5)}$. As in the narrowband biphoton case \cite{1,2,3,4,5,6,7,8,9,10,11,12,13,14,15,16,17,18,19,20,21,22,23}, the triphoton bandwidth typically falls below the GHz range, allowing full resolution by commercial single-photon detectors. Thus, photon-coincidence measurements map directly onto Glauber correlation functions without requiring time averaging. In particular, the threefold coincidence counting rate $R_3$ corresponds to the third-order correlation function $G^{(3)}(t_1,t_2,t_3)$ for photons $\hat{E}_{s1}$, $\hat{E}_{s2}$, and $\hat{E}_{s3}$ jointly detected at the position-time coordinates $(r_1,t_1)$, $(r_2,t_2)$, and $(r_3,t_3)$ of single-photon detectors D$_1$, D$_2$, and D$_3$:
\begin{eqnarray}
R_3=|\langle0|E^{(+)}_3E^{(+)}_2E^{(+)}_1|\Psi\rangle|^2=|A_3(\tau_{32},\tau_{31})|^2,
\end{eqnarray}
where $\tau_{ij}=(t_i-t_j)-(r_i-r_j)/c$ and $A_3(\tau_{32},\tau_{31})$ is the three-photon amplitude (or wavefunction). For the twofold coincidence counting rate $R_2$, one can consider, for example, the joint detection of $\hat{E}_{s2}$ and $\hat{E}_{s3}$ while tracing over the undetected $\hat{E}_{s1}$ photons: 
\begin{eqnarray}
R_2=\sum_{\omega_{s1}}|\langle0|a(\omega_{s1})E^{(+)}_3E^{(+)}_2|\Psi\rangle|^2=\sum_{\omega_{s1}}|A_2(\tau_{32})|^2.
\end{eqnarray}
Here, $A_2(\tau_{32})$ is the reduced three-photon amplitude, obtained after tracing out one photon.

Following standard procedure \cite{30,32,33,34} and algebraic manipulation, the amplitudes take the form
\begin{align}
A_3(\tau_{32},\tau_{31})&=A_{3}\iint d\nu_1d\nu_2\chi^{(5)}\Phi\Big(\frac{\Delta kL}{2}\Big)e^{i(\nu_1\tau_{31}+\nu_2\tau_{32})},\\
A_2(\tau_{32})&=A_2\int d\nu_2\chi^{(5)}\Phi\Big(\frac{\Delta kL}{2}\Big)e^{i\nu_2\tau_{32}},
\end{align}
where $A_3$ and $A_2$ are grouped constants. In the above derivations, the frequencies are approximated as small deviations from central frequencies: $\omega_{sj}=\varpi_{sj}+\nu_j$, with $|\nu_j|\ll\varpi_j$. The $\nu_j$ ranges are constrained by $\chi^{(5)}$, the linear susceptibilities $\chi_{sj}$, and potentially narrowband filters. Energy conservation imposes $\nu_1+\nu_2+\nu_3=0$. Using this, expanding the wavenumbers around the center $K_{sj}$ to first order as $k_{sj}=K_{sj}+\nu_j/v_j$, where $v_j=c/[1+(\varpi_{sj}/2)(d\chi_{sj}/d\nu_j)]$ is the group velocity of the $\hat{E}_{sj}$ photon in the ensemble, gives
\begin{eqnarray}
\Delta k=\frac{2\omega_{21}}{c}+\nu_1\Big(\frac{1}{v_1}-\frac{1}{v_3}\Big)-\nu_2\Big(\frac{1}{v_2}+\frac{1}{v_3}\Big).
\end{eqnarray}
Here we assume $k_p-k_{c1}-k_{c2}+K_{s1}-K_{s2}+K_{s3}=0$. Unlike the biphoton case \cite{1,2,3,4,5,6,7,8,9,10,11,12,13,14,15,16,17,18,19,20,21,22,23}, $A_3(\tau_{32},\tau_{31})$ is a 2D convolution in the $(\tau_{32},\tau_{31})$ domain between the inverse Fourier transforms of $\Phi$ and $\chi^{(5)}$. Consequently, both the triphoton time correlations and the state $|\Psi\rangle$ are highly sensitive to the precise forms of $\chi_{sj}$ and $\chi^{(5)}$. The corresponding correlation function $R_3$ is thus governed by the exact optical response functions involved. To our knowledge, the commonly used perturbation chain rule (PCR) model \cite{24,25,26,27,28,30} is not self-consistent when applied to SSWM, as the derived $\chi^{(5)}$ differs across the three generated photons, despite its frequent use in classical nonlinear optics. This lack of a consistent and accurate theoretical model has hindered a full understanding of SSWM-based triphoton generation and precise comparison between theory and experiment. The purpose of this work is to show that the generalized HE method provides not only an accurate description of field-atom interaction in SSWM, but also self-consistent expressions for $\chi_{sj}$ and $\chi^{(5)}$, paving the way for quantitative experimental validation.

\begin{figure}[t]
\centering
  \includegraphics[width=8.5cm]{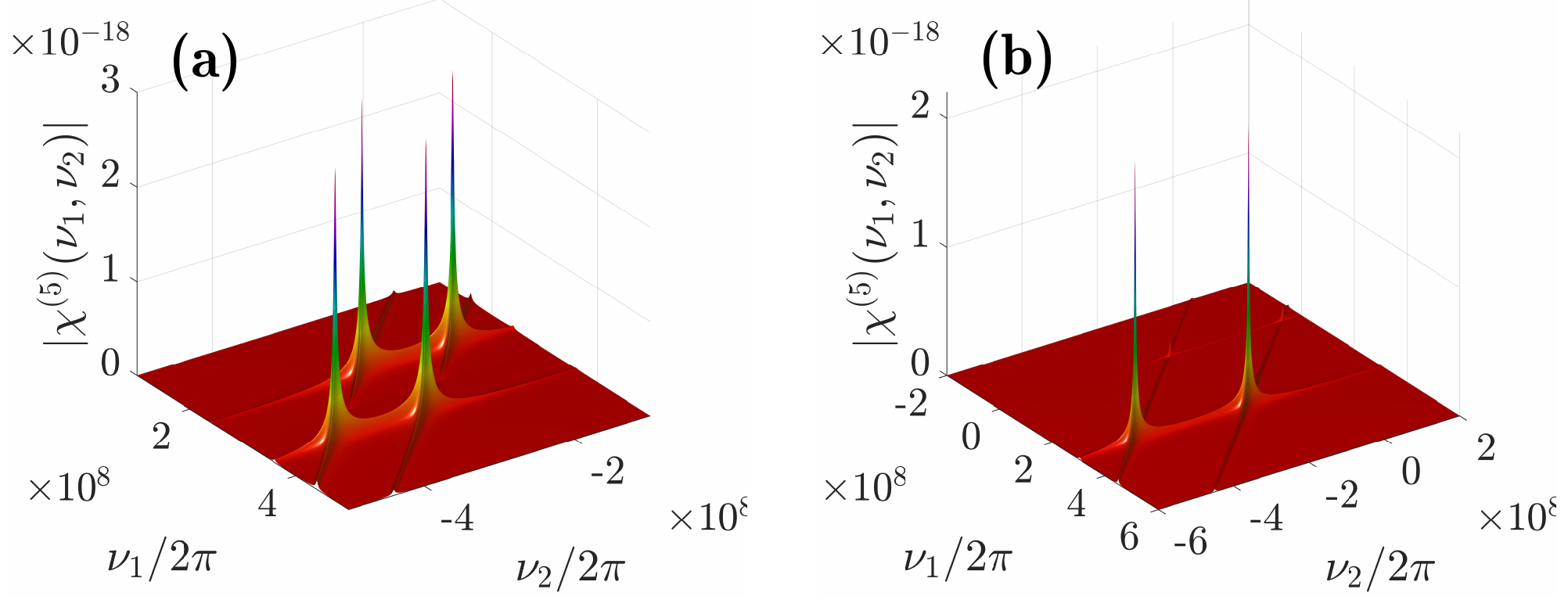}  %\columnwidth
  \caption{Representative resonance profiles of the fifth-order nonlinear susceptibility $|\chi^{5}(\delta_1,\delta_2)|$ for the following parameters: $N=10^{17}$, $L=1.5$ cm, $\gamma_{31}=2\pi\times3$ MHz, $\gamma_{41}=\gamma_{31}$, $\gamma_{51}=0.2\gamma_{31}$, $\gamma_{21}=0.04\gamma_{31}$, $\Delta_{c2}=0$, and $\Delta_p=-2\pi\times300$ MHz. (a) $\Delta_{c1}=0$ MHz, $\Omega_{c1}=\Omega_{c2}=20\gamma_{31}$ and (b) $\Delta_{c1}=300$ MHz, $5\Omega_{c1}=\Omega_{c2}=50\gamma_{31}$.}
  \label{fig2}
\end{figure}

\begin{figure*}[htpb]
\centering
  \includegraphics[width=16cm]{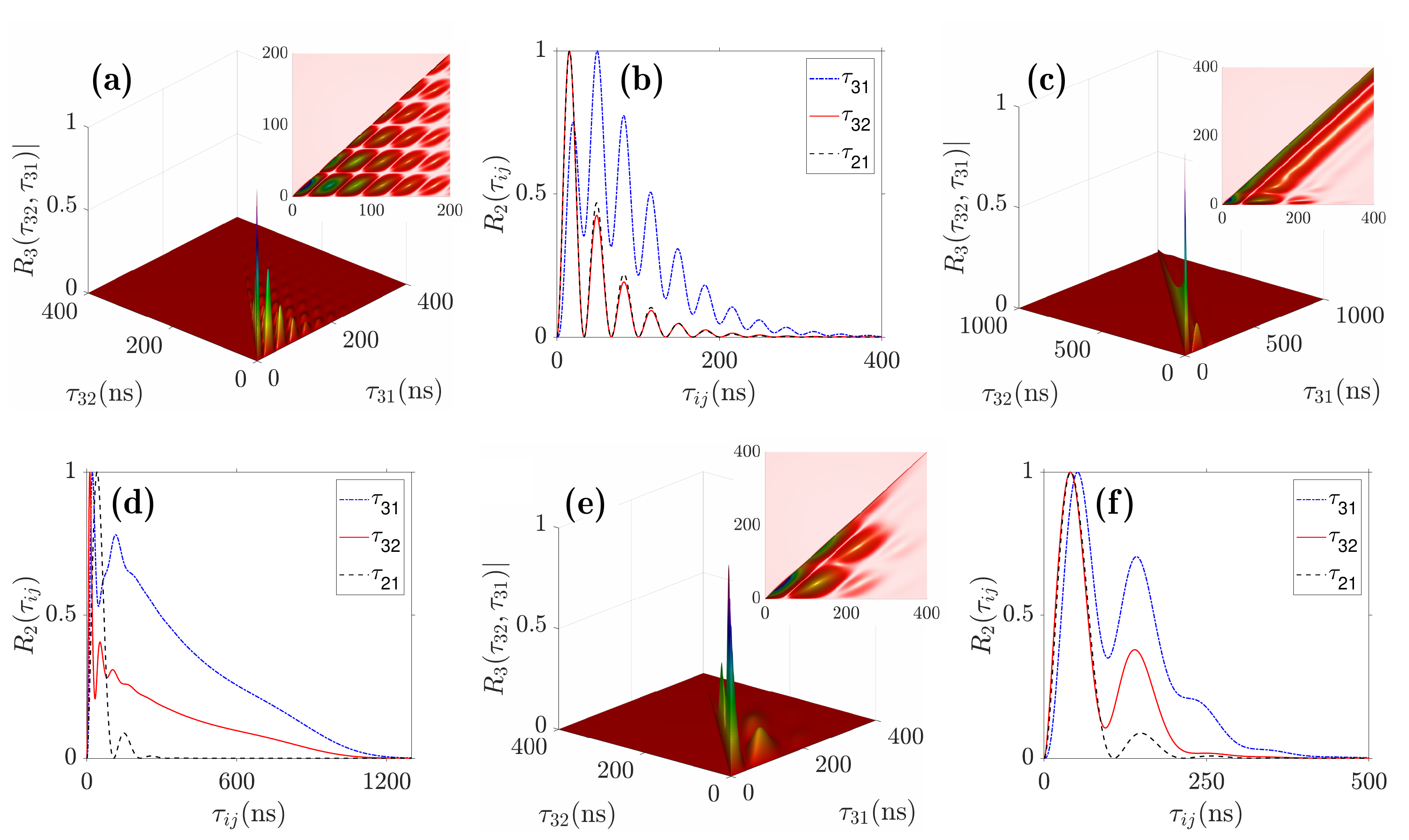}  %\columnwidth
  \caption{Normalized photon coincidence detections $R_3$ and $R_2$ in three operating regimes. (a) Threefold coincidences in the damped Rabi-oscillation regime, using the same parameters as Fig.~\ref{fig2} except $\Omega_{c1}=\Omega_{c2}=5\gamma_{31}$ and $\text{OD}=1.5$; (b) corresponding conditional twofold coincidences. (c) Three-photon coincidences in the group-delay regime, with $\Omega_{c1}=\Omega_{c2}=1.6\gamma_{31}$ and $\text{OD}=88$; (d) corresponding conditional two-photon coincidences. (e) Triple coincidences in the hybrid regime, using parameters from (c) except $\text{OD}=8$; (f) conditional two-photon coincidences in this regime. Insets: top-view contours.}
  \label{fig3}
\end{figure*}

\emph{Generalized HE Theory}---The main theoretical difficulty in finding $\chi_{sj}$ and $\chi^{(5)}$ comes from the need to properly describe the coexistence of the strong coupling field $E_{c1}$ and the generated photon field $\hat{E}_{s1}$ within $|4\rangle\leftrightarrow|5\rangle$ (Fig.~\ref{fig1}(a)). Among existing approaches, the HE method developed by Wen remains uniquely capable of treating such complex light-atom interactions self-consistently. Its accuracy has been well tested in studying narrowband biphoton generation through resonant SFWM across various atomic configurations. 

We extend this method to SSWM triphoton generation depicted in Fig.~\ref{fig1}. Starting from the Heisenberg equations of motion for the atomic operators ($Q_{mn}=|m\rangle\langle n|$) under the dipole approximation \cite{35},
\begin{eqnarray}
\dot{Q}_{mn}=i\omega_{mn}+i\sum^5_{l=1}(d_{nl}\hat{E}\sigma_{ml}-d_{lm}\hat{E}\sigma_{ln}),
\end{eqnarray}
where $d_{mn}=\langle m|\text{d}|n\rangle/\hbar$ is the normalized dipole matrix element and $\hat{E}$ is the total electric-field operator. Its positive-frequency component is $\hat{E}^{(+)}=E_pe^{-i\omega_pt}+E_{c1}e^{-i\omega_{c1}t}+E_{c2}e^{-i\omega_{c2}t}+\sum^3_{j=1}\hat{E}^{(+)}_{sj}e^{-i\omega_{sj}t}$. Applying the rotating-wave approximation and suitable transformations yields a set of slowly varying atomic evolution equations for $\sigma_{mn}$, incorporating phenomenological decay and dephasing rates $\gamma_{mn}$, as detailed in Supplemental Material (SM). These equations (S2) form the basis of the HE methodology. Because they include optical saturation effects, the equations cannot generally be solved analytically. Also, their structure indicates that $\sigma_{mn}$ evolve with contributions from multiple harmonic components. This motivates a harmonic expansion of $\sigma_{mn}$, in which all orders of $E_{c1,c2}$ are retained while only the lowest order in $E_p$ and $\hat{E}_{sj}$ is kept. Thus, the steady-state solutions assume the symmetric HE form \cite{21,22,23}
\begin{eqnarray}
\sigma_{mn}=a_{mn}+b_{mn}e^{-i\nu_1t}+c_{mn}e^{i\nu_1t},
\end{eqnarray}
with $\{|b_{mn}|,|c_{mn}|\}\ll|a_{mn}|$. Since the atomic population is primarily in $|1\rangle$, we set $a_{11}\approx1$. Substituting the trial solution (9) into Eqs.~(S2) and collecting terms with identical harmonic time dependence while neglecting products of small quantities yields, after a lengthy derivation, the following key results:
\begin{align}\label{eq10}
&a_{13}=-\frac{i\digamma^*_{21}\Omega_{s3}^{(+)}}{\digamma^*_{21}\digamma^*_{31}+\left|\Omega_{c2}\right|^2}-\frac{i\Pi^*_{51}\Omega_p\Omega_{c1}\Omega_{c2}\Omega^{(-)}_{s1}\Omega^{(-)}_{s2}}{\Xi^*D^*(\nu_1,\nu_2)}, \nonumber \\
&c_{54}=-\frac{i\left|\Omega_p\Omega_{c1}\right|^2\Omega^{(+)}_{s1}}{\Xi\Gamma_{54}(\Pi_{41}\Pi_{51}+\left|\Omega_{c1}\right|^2)}-\frac{i\Pi_{51}\Omega_p\Omega_{c1}\Omega_{c2}\Omega_{s2}^{(-)}\Omega_{s3}^{(-)}}{\Xi D(\nu_1,\nu_2)},\nonumber \\
&a_{24}=-\frac{i\digamma^*_{53}\left|\Omega_p\Omega_{c1}\right|^2\Omega^{(+)}_{s2}}{\Gamma^*_{54}\Xi^*[\digamma^*_{53}\left|\Omega_{c1}\right|^2+\digamma^*_{42}(\digamma^*_{52}\digamma^*_{53}+\left|\Omega_{c2}\right|^2)]}\nonumber\\
&-\frac{i\Pi^*_{51}\Omega_p\Omega_{c1}\Omega_{c2}\Omega^{(-)}_{s1}\Omega^{(-)}_{s3}}{\Xi^*D^*(\nu_1,\nu_2)},
\end{align}
where $\Omega_j$ denote the Rabi frequencies, and
\begin{align}
D(\nu_1,\nu_2)&=(\Pi_{41}\Pi_{51}+\left|\Omega_{c1}\right|^2)(\digamma_{21}\digamma_{31} +\left|\Omega_{c2}\right|^2),   \nonumber \\
\Xi&=\Gamma_{41}\Gamma_{51}+\left|\Omega_{c1}\right|^2.
\end{align}
Here $\Gamma_{41}=i\Delta_p-\gamma_{41}$, $\Gamma_{51}=i(\Delta_p+\Delta_{c1})-\gamma_{51}$, $\Gamma_{54}=i\Delta_{c1}-\gamma_{54}$, $\digamma_{21}=-i(\nu_1+\nu_2)-\gamma_{21}$, $\digamma_{31}=-i(\nu_1+\nu_2)-\gamma_{31}$, $\digamma_{52}=i(\Delta_p+\Delta_{c1}+\nu_2)-\gamma_{52}$, $\digamma_{53}=i(\Delta_p+\Delta_{c1}+\nu_2)-\gamma_{53}$, $\digamma_{42}=i(\Delta_p+\nu_2)-\gamma_{42}$, and $\Pi_{ij}=\Gamma_{ij}-i\nu_1$. The physics embodied in Eqs.~(10) can be summarized as follows. The first terms on the right-hand side describe the linear optical responses of the created photons $(\hat{E}_{s3},\hat{E}_{s1},\hat{E}_{s2})$, while the second terms correspond to the fifth-order nonlinear response responsible for SSWM. A closer examination shows that the linear responses of $\hat{E}_{s1}$ and $\hat{E}_{s2}$ are negligible, whereas $\hat{E}_{s3}$ photons experience pronounced EIT effects. Moreover, the nonlinear responses are identical for all three channels, strongly confirming the occurrence of genuine triphoton generation via one-step SSWM. These results, together with the susceptibilities derived below, differ substantially from those obtained using the PCR, leading to appreciably different temporal-spectral correlations.

The slowly varying polarization amplitudes of the ensemble are given by $\mathcal{P}_{s3}=N\hbar d_{13}a_{13}$, $\mathcal{P}_{s2}=N\hbar d_{24}a_{24}$, and $\mathcal{P}_{s1}=N\hbar d_{45}c_{45}$. The polarization relates to the respective field through $\mathcal{P}=\epsilon_0\chi E+\epsilon_0\chi^{(5)}E^5$. Plugging Eqs.~(10) into these relations gives
\begin{align}
\chi_{s3}(\nu_1,\nu_2)&=\frac{-iN\hbar|d_{31}|^2\digamma^*_{21}}{\epsilon_0(\digamma^*_{31}\digamma^*_{21}+|\Omega_{c2}|^2)},\\
\chi^{(5)}(\nu_1,\nu_2)&=\frac{-iN\hbar d_{41}d_{32}d_{24}d_{13}|d_{54}|^2\Pi^*_{51}}{\epsilon_0\Xi^*D^*(\nu_1,\nu_2)}.
\end{align}
We neglect $\chi_{s1}$ and $\chi_{s2}$ since their magnitudes scale as $\mathcal{O}(|\Omega_p|^2/|\Omega_{c1,c2}|^2)$, implying that the $\omega_{s1}$ and $\omega_{s2}$ photons sustain negligible gain, loss, and dispersion. In contrast, the $\omega_{s3}$ photons undergo EIT (see SM), characterized by a transparency window $\omega_{\text{tr}}=|\Omega_{c2}|/(\gamma_{31}\sqrt{8\text{OD}})$ and a reduced group velocity $v_3=|\Omega_{c2}|^2L/(\gamma_{31}\text{OD})$, where $\text{OD}=2NL\hbar|d_{31}|^2\varpi_{s3}/(c\epsilon_0\gamma_{31})$ is the optical depth. With these parameters, the longitudinal phase mismatch [Eq.~(7)] becomes $\Delta k=2\omega_{21}/c-(\nu_1+\nu_2)/v_3+i\varpi_{s3}\text{Im}[\chi_{s3}]/c$, where the final term accounts for EIT-induced absorption. As the propagation layer, the narrower of the two spectral widths--between the EIT window $\omega_{\text{tr}}$ and that defined by $\Phi(\Delta kL/2)$--limits the frequency range of triphotons emerging from the medium.

On the other hand, the spectral profile of $\chi^{(5)}$ not only reveals the triphoton generation mechanism through its resonance structures but also defines the intrinsic generation bandwidths determined by the dressed-state effects. Such properties are dictated by the real and imaginary parts of $D(\nu_1,\nu_2)$ \cite{21,22,23}. Though the two-photon resonance condition $\Delta_p+\Delta_{c1}=0$ may enlarge SSWM efficiency, it renders analytical treatment cumbersome. To clearly illustrate the generation mechanism and distinguish SSWM from cascaded SFWMs \cite{33}, we set $\Delta_{c1}\to0$ for simplicity. Despite the dependence of $D(\nu_1,\nu_2)$ on two independent frequency variables, its real-part roots exhibit a clear resonance pattern. Along the $\nu_1$ axis, two resonances appear at $\Delta_{p}\pm\Omega_{e1}/2$ with linewidth $\gamma_{e1}$, where $\Omega_{e1}=\sqrt{4|\Omega_{c1}|^2-(\gamma_{41}-\gamma_{51})^2}$ and $\gamma_{e1}=(\gamma_{41}+\gamma_{51})/2$. Along the $\nu_2$ axis, four resonances emerge at $-\Delta_p\pm\Omega_{e1}/2\pm\Omega_{e2}/2$, each with linewidth $\gamma_{e2}$, where $\Omega_{e2}=\sqrt{4|\Omega_{c2}|^2-(\gamma_{31}-\gamma_{21})^2})$ and $\gamma_{e2}=(\gamma_{21}+\gamma_{31})/2$. These features suggest four distinct yet mutually coherent SSWM pathways that interfere both constructively and destructively in the temporal domain. The corresponding triphoton resonances occur at: (i) $(\omega_{s1},\omega_{s2},\omega_{s3})=(\varpi_{s1}-\Delta_{p}+\Omega_{e1}/2,\varpi_{s2}+\Delta_{p}-\Omega_{e1}/2+\Omega_{e2}/2,\varpi_{s3}-\Omega_{e2}/2)$; (ii) $(\varpi_{s1}-\Delta_{p}+\Omega_{e1}/2,\varpi_{s2}+\Delta_{p}-\Omega_{e1}/2-\Omega_{e2}/2,\varpi_{s3}+\Omega_{e2}/2)$; (iii) $(\varpi_{s1}-\Delta_{p}-\Omega_{e1}/2,\varpi_{s2}+\Delta_{p}+\Omega_{e1}/2+\Omega_{e2}/2,\varpi_{s3}-\Omega_{e2}/2)$; and $(\varpi_{s1}-\Delta_{p}-\Omega_{e1}/2,\varpi_{s2}+\Delta_{p}+\Omega_{e1}/2-\Omega_{e2}/2,\varpi_{s3}+\Omega_{e2}/2)$. All four configurations naturally satisfy energy conservation, confirming their consistency with the underlying parametric processes. A representative $|\chi^{(5)}|$ spectrum using experimentally accessible parameters is plotted in Fig.~\ref{fig2}(a), where four well-resolved resonant peaks are evident. Moreover, energy conservation $\nu_1+\nu_2+\nu_3=0$ requires $|\chi^{(5)}|$ to exhibit central symmetry about its global center. Notably, when $\Delta_{c1}$ and $\Delta_{c2}$ deviate from zero and $|\Omega_{c1}|$ and $|\Omega_{c2}|$ are largely unequal, the resonance structures can change markedly--the peaks may vary in height, and the number of observable resonances may decrease (Fig.~\ref{fig2}(b)). For reference, we compare $\chi_{sj}$ and $\chi^{(5)}$ with the results attained from the PCR approach in the SM.

The bandwidths arising from $\Delta\omega_{\text{tr}}$ and $\Phi$ due to propagation collectively interplay with the emission linewidths $\gamma_{e1}$ and $\gamma_{e2}$ to shape the triphoton spectral correlations, which in turn tailor the temporal behavior observed in coincidence measurements. Interestingly, $\chi_{s3}$ can be visualized as a 2D surface in the $\omega_{s3}$ photon's own spectral space, but as a 3D structure when represented in the joint $(\omega_{s1},\omega_{s2})$ space--an intrinsic manifestation of the continuous-mode W-state property \cite{36} that has no analogue in the biphoton regime. In contrast, $\chi^{(5)}$ always has a 3D spectral distribution, reflecting the full tripartite entanglement regardless of which two-photon subspace is considered. The interplay among these spectral widths physically defines two distinct regimes of triphoton temporal correlations. However, a broad intermediate region also exists, where one temporal dimension is dominated by propagation-induced effects while the other is governed by the spectral width of $\chi^{(5)}$--a feature also unique to genuine triphoton systems. To elucidate these regimes, we next analyze temporal correlation characteristics.

\emph{Temporal Correlations}---With Eqs.~(12) and (13), we are now ready to examine the temporal correlation properties of the W-triphoton state through the third- and second-order correlation functions [Eqs.~(3) and (4)]. In general, obtaining analytical solutions is challenging. However, when either $\Phi$ or $\chi^{(5)}$ dominates the generation process, closed-form solutions can be derived. These two limiting cases--analogous to biphoton generation--correspond to the {\it group-delay regime} and {\it damped Rabi oscillation regime}, respectively. Beyond these limits, analytical treatment becomes intractable, requiring numerical analysis. To illustrate the distinctions among these regions, we now examine each case in detail.

\underline{(I) Damped Rabi Oscillations.} In this regime, $\chi^{(5)}$ essentially governs the spectral bandwidths over which triphotons are output from the ensemble. The sinc term in $\Phi$ can be approximated as unity, though its residual phase factor remains important for shaping temporal correlations. While an analytical solution could, in principle, be derived via the Cauchy residue theorem--as done for biphotons \cite{21,22,23}--the resulting formula is lengthy and offers limited physical transparency. Because of the finite timing resolution of commercial single-photon detectors, the measured temporal correlations represent a complex superposition of indistinguishable SSWM pathways. To elucidate the essential dynamics, we present numerical results in Figs.~\ref{fig3}(a) and (b), showing $R_3$ and $R_2$, respectively. The four SSWM channels interfere both constructively and destructively, producing a striking 3D temporal structure with maximal coherence along the $\tau_{31}+\tau_{32}$ diagonal. The inset contour clearly reveals that oscillations are confined below this diagonal, while the upper region is excluded by the energy-conservation constraint inherent to W-triphoton formation. Furthermore, the conditional two-photon correlations--obtained by tracing over any single photon--exhibit damped Rabi oscillations similar to those observed in SFWM-based biphoton systems, except that their periods are determined by nontrivial combinations $\Omega_{e1}$ and $\Omega_{e2}$. 

\underline{(II) Group Delay.} Unlike the damped Rabi-oscillation regime, here $\Phi$ becomes the primary factor determining the spectral bandwidth over which triphotons propagate through the medium at large optical depth. The effective bandwidth is set by the narrower of two limits: the sinc-induced width $\Delta\omega_{\text{sl}}$ and the EIT transmission window $\Delta\omega_{\text{tr}}$. Since the linewidths $\gamma_{e1}$ and $\gamma_{e2}$ associated with the $\chi^{(5)}$ resonances are comparatively broad, $\chi^{(5)}$ may be treated as approximately constant. In the ideal limit where $\Delta\omega_{\text{sl}}\ll\Delta\omega_{\text{tr}}$, the threefold coincidence distribution is expected to be a nearly square profile below the $\tau_{31}=\tau_{32}$ diagonal, free of oscillations. When $\Delta\omega_{\text{tr}}<\Delta\omega_{\text{sl}}$, the pattern evolves into a decaying, square-like structure. In practice, residual $\chi^{(5)}$ contributions remain unavoidable, slightly modulating the temporal correlations. As illustrated in Figs.~\ref{fig3}(c) and (d), the pronounced EIT experienced by the $\omega_{s3}$ photons surpasses the damped Rabi oscillations observed previously. The resulting decay dynamics are governed mostly by EIT-induced losses, which dramatically extend the coherence time, as reflected in both $R_3$ and the conditional correlations $R_2(\tau_{31})$ and $R_2(\tau_{32})$. 

\underline{(III) Hybrid Regime.} Between the two limits described above lies a broad intermediate region where both the phase-mismatch function $\Phi$ and the fifth-order susceptibility $\chi^{(5)}$ contribute comparably to manipulating the temporal correlations. In this mixed regime, analytical simplifications are no longer sufficient, and the threefold coincidence rate $R_3$ together with the conditional twofold rates $R_2$ must be evaluated numerically. As shown in Figs.~\ref{fig3}(e) and (f), the resulting correlation patterns exhibit features characteristic of both the damped Rabi-oscillation and group-delay regimes, reflecting a gradual transition between them.

In summary, we have generalized the HE framework to provide an accurate and self-consistent description of continuous-mode, time-energy-entangled W-triphotons generated resonantly through the SSWM process in a five-level asymmetric-M atomic system confined within a 2D magneto-optical trap. This approach captures both linear and fifth-order nonlinear optical responses within a unified formalism, offering clear advantages over the PCR technique, which treats these effects qualitatively. The resulting predictions--together with the explicit comparisons presented here--lay a solid foundation for forthcoming experimental tests and for exploring higher-order quantum correlations in intricate atomic media. 

We gratefully acknowledge support from Binghamton University through startup funding.

\end{document}

\title{Supplemental Material for ``Triphoton generation near atomic resonance via SSWM: Harmonic expansion for accurate optical response''}

% Use the exact author list and affiliations from the main text, if required by the journal.
\author{Jianming Wen}
\affiliation{Department of Electrical and Computer Engineering, Binghamton University, Binghamton, New York 13902, USA}
% ...

\date{\today}

% A brief opening paragraph is optional; PRL typically prefers direct sections.
\maketitle

% ---------- Outline (example) ----------
% 1) Experimental details
% 2) Additional derivations
% 3) Supplemental figures & tables
% 4) Data analysis & uncertainty
% 5) Additional references (longbibliography)

% ---------- Section 1: Additional Theory/Derivations ----------
\section{Additional Theory and Derivations}
To remove the rapidly oscillating phase factors in Eqs.~(8) of the main text, we have introduced the following transformations:
\begin{align}
\sigma_{jj}&=Q_{jj},\quad \sigma_{21}=Q_{21}e^{-i(\omega_{s3}-\omega_{c2})t},\quad \sigma_{31}=Q_{31}e^{-i\omega_{s3}t},\quad 
\sigma_{32}=Q_{32}e^{-i\omega_{c2}t},\nonumber\\
\sigma_{41}&=Q_{41}e^{-i\omega_pt},\quad \sigma_{42}=Q_{42}e^{-i\omega_{s2}t},\quad \sigma_{43}=Q_{43}e^{-i(\omega_{s2}-\omega_{c2})t},\quad
\sigma_{51}=Q_{51}e^{-i(\omega_p-\omega_{c1})t},\nonumber\\
\sigma_{52}&=Q_{52}e^{-i(\omega_{s2}+\omega_{c1})t},\quad
\sigma_{53}=Q_{53}e^{-i(\omega_{s2}+\omega_{c1}-\omega_{c2})t},\quad
\sigma_{54}=Q_{54}e^{-i\omega_{c1}t}.
\end{align}
Plugging Eqs.~(S1) into Eqs.~(8) and applying the rotating-wave approximation gives a set of coupled equations that goven the atomic evolution:
\begin{align}
\dot{\sigma}_{21}&=\digamma_{21}\sigma_{21}+id_{13}\hat{E}_{s3}^{(-)}\sigma_{23}-id_{32}E_{c2}\sigma_{31}+id_{14}E_p^*\sigma_{24}e^{-i\nu_1t}-id_{42}\hat{E}_{s2}^{(+)}\sigma_{41}e^{-i\nu_1t},   \nonumber \\
\dot{\sigma}_{31}&=\digamma_{31}\sigma_{31}-id_{13}\hat{E}_{s3}^{(-)}-id_{23}E^*_{c2}\sigma_{21}+id_{14}E_p^*\sigma_{34}e^{-i\nu_1t}, \nonumber \\
\dot{\sigma}_{32}&=\Gamma_{32}\sigma_{32}-id_{13}\hat{E}_{s3}^{(-)}\sigma_{12}+id_{24}\hat{E}_{s2}^{(-)}\sigma_{34},        \nonumber \\
\dot{\sigma}_{41}&=\Gamma_{41}\sigma_{41}-id_{14}E_p^*-id_{24}\hat{E}_{s2}^{(-)}\sigma_{21}e^{i\nu_1t}+id_{13}\hat{E}_{s3}^{(-)}\sigma_{43}e^{i\nu_1t}-id_{54}[E_{c1}+\hat{E}_{s1}^{(+)}e^{i\nu_1t}]\sigma_{51},  \nonumber \\
\dot{\sigma}_{42}&=\digamma_{42}\sigma_{42}-id_{14}E_p^*\sigma_{12}e^{-i\nu_1t}+id_{23}E^*_{c2}\sigma_{43}-id_{54}[E_{c1}+\hat{E}_{s1}^{(+)}e^{i\nu_1t}]\sigma_{52},\nonumber\\
\dot{\sigma}_{43}&=\digamma_{43}\sigma_{43}+id_{31}\hat{E}_{s3}^{(+)}\sigma_{41}e^{-i\nu_1t}-id_{14}E_p^*\sigma_{13}e^{-i\nu_1t}+id_{32}E_{c2}\sigma_{42}-id_{24}\hat{E}_{s2}^{(-)}\sigma_{23}-id_{54}[E_{c1}+\hat{E}_{s1}^{(+)}e^{i\nu_1t}]\sigma_{53},   \nonumber \\
\dot{\sigma}_{51}&=\Gamma_{51}\sigma_{51}+id_{13}\hat{E}_{s3}^{(-)}\sigma_{53}e^{i\nu_1t}+id_{14}E_p^*\sigma_{54}-id_{45}[E^*_{c1}+\hat{E}_{s1}^{(-)}e^{-i\nu_1t}]\sigma_{41},      \nonumber \\
\dot{\sigma}_{52}&=\digamma_{52}\sigma_{52}+id_{23}E^*_{c2}\sigma_{53}+id_{24}\hat{E}_{s2}^{(-)}\sigma_{54}-id_{45}[E^*_{c1}+\hat{E}_{s1}^{(-)}e^{-i\nu_1t}]\sigma_{42},  \nonumber \\
\dot{\sigma}_{53}&=\digamma_{53}\sigma_{53}+id_{31}\hat{E}_{s3}^{(+)}\sigma_{51}e^{-i\nu_1t}+id_{32}E_{c2}\sigma_{52}-id_{45}[E^*_{c1}+\hat{E}_{s1}^{(-)}e^{-i\nu_1t}]\sigma_{43}, \nonumber \\
\dot{\sigma}_{54}&=\Gamma_{54}\sigma_{54}+id_{41}E_p\sigma_{51}+id_{42}\hat{E}_{s2}^{(+)}\sigma_{52}.
\end{align}
Here we have defined
\begin{align}
\nu_1&=\omega_{c1}-\omega_{s1},\quad\nu_2=\omega_{42}-\omega_{s2},\quad
\digamma_{21}=-i(\nu_1+\nu_2)-\gamma_{21},\quad\digamma_{31}=i(\Delta_{c2}-\nu_1-\nu_2)-\gamma_{31},\quad\Gamma_{32}=i\Delta_{c2}-\gamma_{32},\nonumber\\
\Gamma_{41}&=i\Delta_p-\gamma_{41},\quad\digamma_{42}=i(\Delta_p+\nu_2)-\gamma_{42},\quad\digamma_{43}=i(\nu_2+\Delta_p-\Delta_{c2})-\gamma_{43},\quad\Gamma_{51}=i(\Delta_p+\Delta_{c1})-\gamma_{51},\nonumber\\
\digamma_{52}&=i(\nu_2+\Delta_p+\Delta_{c1})-\gamma_{52},\quad
\digamma_{53}=i(\nu_2+\Delta_p+\Delta_{c1}-\Delta_{c2})-\gamma_{53},\quad\Gamma_{54}=i\Delta_{c1}-\gamma_{54},
\end{align}
where the parameters $\gamma_{mn}$ are the dephasing rates between levels $|m\rangle$ and $|n\rangle$. The principal theoretical difficulty in solving Eqs.~(S2) stems from the persistence of slowly-varying terms $e^{\pm i\nu_1t}$, together with the nonlinear interplay of the strong coupling laser field E$_{c1}$ and the quantized signal field $\hat{E}_{s1}$ in the $|4\rangle\leftrightarrow|5\rangle$ channel. From Eqs.~(S2), we define the Rabi frequencies $\Omega_p=d_{41}E_p$ for the weak pump laser, $\Omega_{c1}=d_{54}E_{c1}$ and $\Omega_{c2}=d_{32}E_{c2}$ for the two coupling lasers, and $\Omega^{(+)}_{s1}=d_{54}\hat{E}^{(+)}_{s1}$, $\Omega^{(+)}_{s2}=d_{42}\hat{E}^{(+)}_{s2}$, and $\Omega^{(+)}_{s3}=d_{31}\hat{E}^{(+)}_{s3}$ for the $\omega_{s1}$, $\omega_{s2}$, and $\omega_{s3}$ photon fields, respectively, in order to simplify the notation used in Eqs.~(10) of the main text.

%----------Section 2: Further Disussion on Linear Susceptibility------
\section{Further Discussion on the Linear Susceptibility $\chi_{s3}(\nu_1,\nu_2)$}
Different from the biphoton case---and absent in classical EIT studies---the direct generation of genuine W-triphotons offers a spectacular perspective on the standard EIT effect undergone by the $\omega_{s3}$ photons through the correlations with the $\omega_{s1}$ and $\omega_{s2}$ photons. This interesting connection arises naturally from their shared energy conservation condition, $\nu_1+\nu_2+\nu_3=0$. 

\begin{figure}[htpb]
  \centering
  \includegraphics[width=0.75\linewidth]{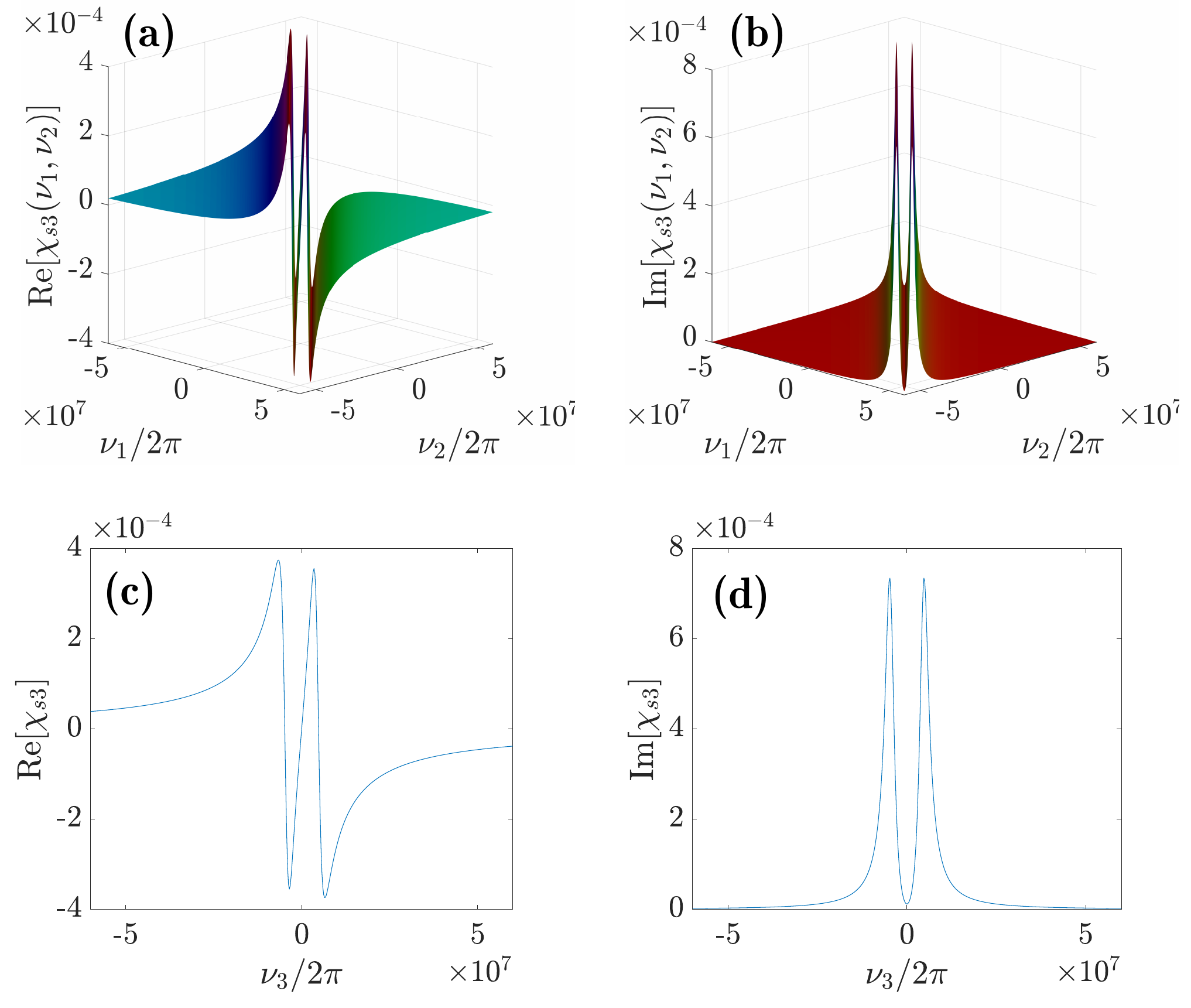}
  \caption{3D spectral representation of the linear susceptibility $\chi_{s3}(\nu_1,\nu_3)$ viewed from the frequency domains of the correlated $\omega_{s1}$ and $\omega_{s2}$ photons. (a) Real component, showing a 3D dispersive spectral surface. (b) Imaginary component, exhibiting an absorptive 3D spectral structure. For comparison, panels (c) and (d) display the EIT characteristics of $\chi_{s3}(\nu_3)$ directly in the spectral domain of the $\omega_{s3}$ photons. Owing to the W-type temporal-energy entanglement, the sharp dispersion and transparency signatures associated with $\omega_{s3}$ can be indirectly perceived through the $\omega_{s1}$ and $\omega_{s2}$ channels.}
  \label{figS1}
\end{figure}

In the $\omega_{s3}$ photon's own frame, the real and imaginary parts of $\chi_{s3}(\nu_3=-\nu_1-\nu_2)$ showcase the familiar line shapes of a standard three-level $\Lambda$-type EIT system; see Figs.~\ref{figS1}(c) and (d). However, when $\chi_{s3}(\nu_1,\nu_2)$ is visualized in the frequency spaces of the $\omega_{s1}$ and $\omega_{s2}$ photons, its real and imaginary components form three-dimensional (3D) nontrivial dispersive and absorptive spectral landscapes as illustrated in Figs.~\ref{figS1}(a) and (b), respectively. The EIT features, as expected, become more pronounced along the $\nu_1=\nu_2$ direction. Remarkably, when viewed from only either the $\omega_{s1}$ or $\omega_{s2}$ photon's frame, the characteristic EIT features of the $\omega_{s3}$ photons are effectively {\it replicated} to some extent---even through the $\omega_{s1}$ and $\omega_{s2}$ photons themselves do not experience EIT directly. This phenomenon stems from the intrinsic W-type entanglement in time and energy among the three photons. As such, Figs.~\ref{figS1}(a)-(d) intuitively illustrate these correlations and their manifestations in the spectral domain.

%--------Section 3: Further Discussion on the fifth-order susceptibility------------
\section{Further Discussion on the Fifth-Order Nonlinear Susceptibility $\chi^{(5)}(\nu_1,\nu_2)$}
As discussed in the main text, the resonance structure of the fifth-order nonlinear susceptibility $\chi^{(5)}(\nu_1,\nu_2)$ is governed by the interplay between the frequency detunings $\Delta_p,\Delta_{c1},\Delta_{c2}$ and the coupling Rabi frequencies $\Omega_{c1},\Omega_{c2}$. This interplay modulates the dressing strength of the atomic transitions, thereby altering both the number and the relative amplitudes of the resonances. The number of resonant peaks directly reflects how many distinct yet mutually coherent SSWM pathways contribute to the direct generation of W-triphotons.

Because commercial single-photon detectors possess detection bandwidths far broader than the spectral ranges defined by either $\chi^{(5)}$ or $\Phi$, all generated narrowband triphotons are registered indistinguishably. The resulting superposition produces constructive and destructive 3D nonlocal interference patterns, observable in both three-photon ($R_3$) and conditional two-photon ($R_2$) temporal correlations.

\begin{figure*}[htpb]
\centering
  \includegraphics[width=17cm]{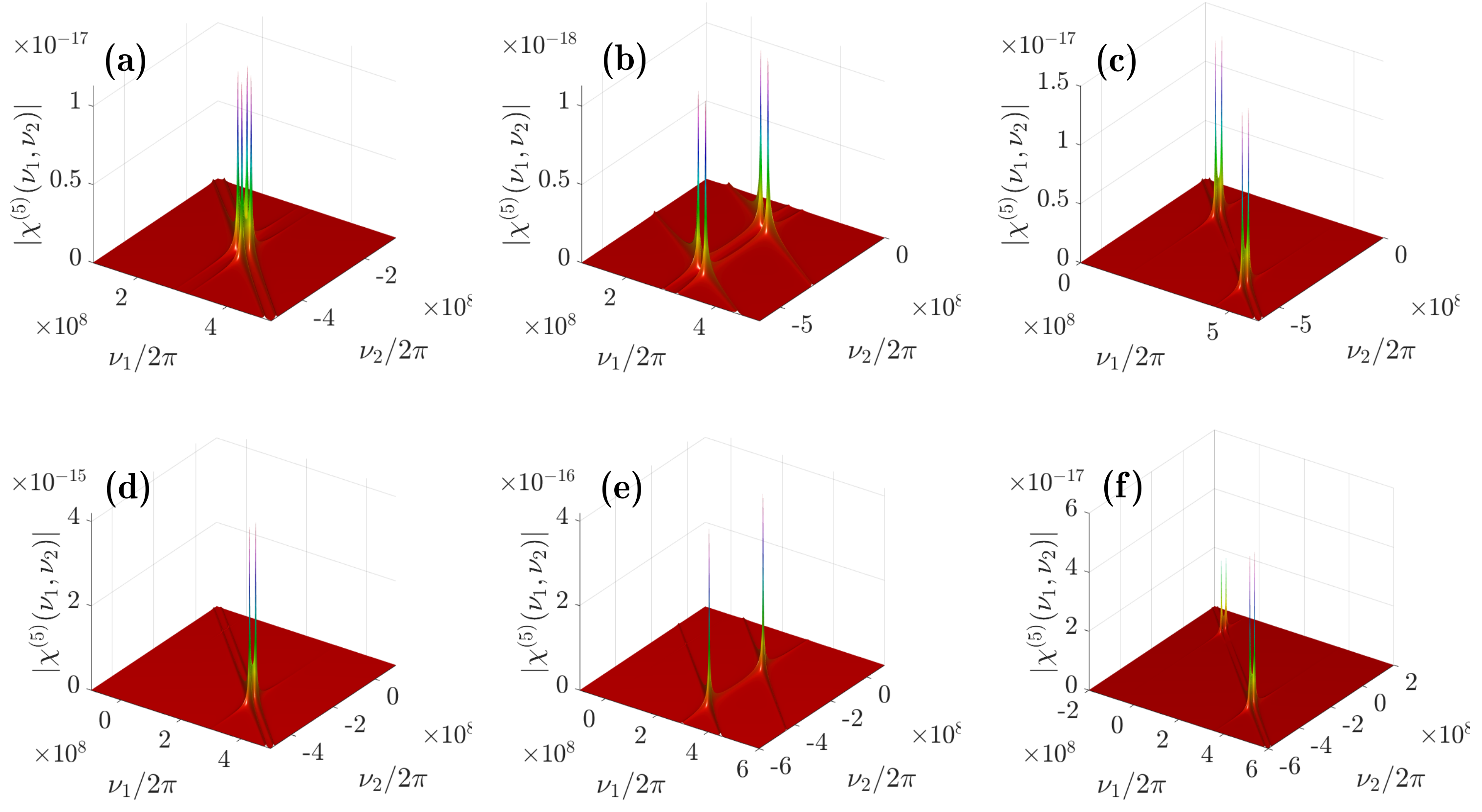}  %\columnwidth
  \caption{Exemplary 3D spectral profiles of the fifth-order nonlinear susceptibility $|\chi^{5}(\nu_1,\nu_2)|$. Panels (a)-(c) correspond to $\Omega_{c1}=\Omega_{c2}=5\gamma_{31}$, $10\Omega_{c1}=\Omega_{c2}=50\gamma_{31}$, and $\Omega_{c1}=10\Omega_{c2}=50\gamma_{31}$, respectively, with all other parameters same as those in Fig.~2(a) of the main text. Panels (d)-(f) use the same conditions as (a)-(c) except for a nonzero detuning $\Delta_{c1}=300$ MHz.}
  \label{figS2}
\end{figure*}

Moreover, by adjusting system parameters such as $\Delta_{c1},\Delta_{c2},\Omega_{c1}$, and $\Omega_{c2}$, one can tune the relative resonance amplitudes. When these resonances acquire unequal heights, the produced W-triphoton state becomes non-maximally entangled, and the constructive and destructive interferences no longer fully cancel. Figures~\ref{figS2}(a)-(f) illustrate representative examples of $|\chi^{(5)}(\nu_1,\nu_2)|$ under various parameter settings: panels (a)-(c) show four resonances of equal height that can be spectrally shifted by tuning the control laser powers, whereas (d) and (e) demonstrate cases with fewer resonances. Introducing a nonzero detuning $\Delta_{c1}$ (panel (f)) further leads to asymmetrical resonance amplitudes.

These results hence highlight how parameter-driven complexity not only enriches the temporal-correlation behavior of W-triphotons but also raises new theoretical and experimental challenges in advancing our understanding of triphoton optics.

%-------------Section 4: Comparison with the PCR Results-------------
\section{Comparison with the Perturbation Chain Rule Method}
Our extended HE method predicts that only the $\omega_{s3}$ photons are subject to a non-negligible linear optical response, and all newborn photons share identical fifth-order nonlinear optical responses. To further substantiate our findings, it is instructive and illuminating to directly compare the HE results with those obtained using the perturbation chain rule (PCR) method. 

For the same five-level asymmetric-M atomic configuration depicted in Fig.~1(a) of the main text, and under the same geometrical arrangement of field-atom interactions given in Fig.~1(b), the PCR approach typically proceeds through the following perturbative chains to reach the relevant density matrix elements:
\begin{align}
\rho_{11}\stackrel{\omega_p}{\longrightarrow}\rho_{41}\stackrel{\omega_{c1}}{\longrightarrow}\rho_{51}\stackrel{\omega_{s1}}{\longrightarrow}\rho_{41}\stackrel{\omega_{s2}}{\longrightarrow}\rho_{21}\stackrel{\omega_{c2}}{\longrightarrow}\rho_{31}; \\
\rho_{22}\stackrel{\omega_{c2}}{\longrightarrow}\rho_{32}\stackrel{\omega_{s3}}{\longrightarrow}\rho_{12}\stackrel{\omega_{p}}{\longrightarrow}\rho_{42}\stackrel{\omega_{c1}}{\longrightarrow}\rho_{52}\stackrel{\omega_{s1}}{\longrightarrow}\rho_{42}; \\
\rho_{44}\stackrel{\omega_{s2}}{\longrightarrow}\rho_{24}\stackrel{\omega_{c2}}{\longrightarrow}\rho_{34}\stackrel{\omega_{s3}}{\longrightarrow}\rho_{14}\stackrel{\omega_{p}}{\longrightarrow}\rho_{44}\stackrel{\omega_{c1}}{\longrightarrow}\rho_{54}.
\end{align}
From these density matrix elements, one can arrive at the corresponding expressions for the linear susceptibilities $\chi_{sj}$,
\begin{align}
&\chi_{s1}(\nu_1)=\frac{-iN\hbar|d_{54}|^2(\nu_1+i\gamma_{54})}{\varepsilon_0\left.[(\Delta_{c1}+\nu_1+i\gamma_{54})(\nu_1+i\gamma_{54})+\left|\Omega_{c1}\right|^2\right.]}, \\
&\chi_{s2}(\nu_2)=\frac{-iN\hbar|d_{42}|^2(\Delta_{p}+\nu_2+i\gamma_{32})}{\varepsilon_0\left.[(\Delta_{p}+\nu_2+i\gamma_{42})(\Delta_{p}+\nu_2+i\gamma_{32})+\left|\Omega_{c2}\right|^2\right.]}, \\
&\chi_{s3}(\nu_3)=\frac{-iN\hbar|d_{31}|^2\digamma_{21}}{\varepsilon_0(\digamma_{31}\digamma_{21}+\left|\Omega_{c2}\right|^2)},
\end{align}
and the fifth-order nonlinear susceptibilities $\chi^{(5)}_{sj}$,
\begin{align}
\chi^{(5)}_{s3}=\frac{-iN\hbar d_{41}d_{32}d_{24}d_{13}|d_{54}|^2}{\varepsilon_0\left.(\Gamma_{41}\Gamma_{51}+\left|\Omega_{c1}\right|^2\right.)(i\Delta_p-i\nu_1-\gamma_{41})(\digamma_{21}\digamma_{31} +\left|\Omega_{c2}\right|^2)}
\end{align}
through the $\omega_{s3}$ photon generation chain (S4), 
\begin{align}
\chi^{(5)}_{s2}=\frac{-iN\hbar d_{41}d_{32}d_{24}d_{13}|d_{54}|^2\gamma_{22}}{\varepsilon_0\Pi(i\nu_1+i\nu_2-\gamma_{21})\left.[(i\Delta_p+i\nu_1+i\nu_2-\gamma_{41})(i\Delta_p+i\Delta_{c1}+i\nu_1+i\nu_2-\gamma_{51})+\left|\Omega_{c1}\right|^2\right.]
(i\Delta_p+i\nu_2-\gamma_{41})}
\end{align}
utilizing the $\omega_{s2}$ photon generation chain (S5), with $\Pi=\gamma_{32}\gamma_{22}+\left|\Omega_{c2}\right|^2$, and
\begin{align}
\chi^{(5)}_{s1}&=\frac{-iN\hbar d_{41}d_{32}d_{24}d_{13}|d_{54}|^2}{\varepsilon_0\left.[(-i\Delta_p-i\nu_2-\gamma_{42})(-i\Delta_p-i\nu_2-\gamma_{43})+\left|\Omega_{c2}\right|^2\right.](-i\Delta_p+i\nu_1-\gamma_{41})\left.[(i\nu_1-\gamma_{44})(i\nu_1+i\Delta_{c1}-\gamma_{54})+\left|\Omega_{c1}\right|^2\right.]},
\end{align}
according to the $\omega_{s1}$ photon generation chain (S6).

By comparing with the HE results [Eqs.~(12) and (13)] presented in the main text, we immediately see that, for the linear susceptibilities, the HE and PCR methods yield the only identical result for $\chi_{s3}$ but exhibit clear discrepancies for $\chi_{s1}$ and $\chi_{s2}$. More notably, they completely disagree on $\chi^{(5)}$, which constitutes the essential physical mechanism driving W-triphoton generation in the SSWM process. Our HE method provides a self-consistent expression for $\chi^{(5)}$, independent of which photon is taken as the starting point, as long as they originate from the same SSWM channel. In sharp contrast, the PCR approach produces three distinct fifth-order susceptibilities [Eqs.~(S10)–(S12)], none of which match the HE result. This divergence further underscores the internal inconsistency of the PCR method when extending to the spontaneous multi-wave mixing domain. 

%-----------Section5: Temporal correlation-------------------
\section{Triphoton Temporal Correlations with Two Sets of SSWMs}
\begin{figure}
    \centering
    \includegraphics[width=1.0\linewidth]{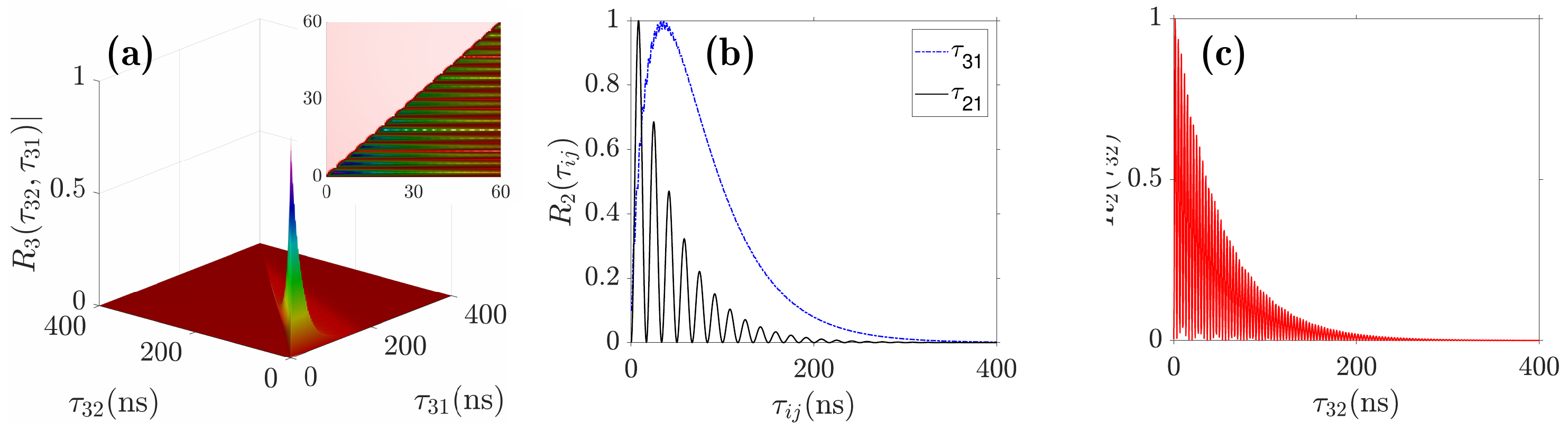}
    \caption{Triphoton temporal correlations where $\chi^{(5)}(\nu_1,\nu_2)$ features two resonances. (a) Normalized three-photon coincidence counting rate and inset: top-view contour; (b), (c) conditional two-photon coincidence counting rates. Parameters are same as those used in Fig.~2(b) of the main text.}
    \label{figS3}
\end{figure}

Figure~2(b) in the main text, together with Figs.~\ref{figS2}(d) and (e) above, explicitly illustrates a case where the fifth-order susceptibility $\chi^{(5)}(\nu_1,\nu_2)$ displays only two distinct resonances, achieved by engineering the dressing effects through adjustments of the coupling frequency detunings and Rabi frequencies. In contrast to the four-resonance scenario, the triphoton temporal correlation patterns change dramatically when only two sets of SSWM trajectories contribute. To explore the corresponding temporal behavior, Figs.~\ref{figS3}(a)-(c) present numerically simulated triphoton and conditional biphoton coincidence rates. As we can see from Fig.~\ref{figS3}(a), although the three-photon joint detection rate remains confined to the left-right subspace defined by $\tau_{31}=\tau_{32}$, the pattern differs markedly from that of the four-pathway situation [Figs.~3(a), (c), and (e) in the main text]. In this two-resonance configuration, constructive and destructive interferences occur along distinct, well-defined directions (see the inset contour). These modifications are further reflected in the conditional two-photon temporal correlations shown in Figs.~\ref{figS3}(b) and (c), where the residual two-photon temporal entanglement approaches the characteristics of SFWM-generated photon pairs. A more detailed investigation of this special case would be a worthwhile subject for future study.

% ---------- References ----------
% Use the same .bib file as the main article or a separate one. APS style uses apsrev4-2.
%\bibliographystyle{apsrev4-2}
% If using a separate bibliography for the supplement, replace `main` with `supplement` or desired .bib name.
%\bibliography{main}

% ---------- End ----------